%% file: fiz-template.tex
\documentclass{fizik}
\usepackage{hyperref}
\hypersetup{
colorlinks=true,
urlcolor=blue,
citecolor=blue}
\usepackage[all]{xy,xypic}
\usepackage{amsfonts,amssymb,amsmath,amsgen,amsopn,amsbsy,theorem,graphicx,epsfig}
\usepackage{eufrak,amscd,bezier,latexsym,mathrsfs,enumerate,multirow}
\usepackage[utf8]{inputenc}\usepackage[english]{babel}
\usepackage[dvipsnames]{xcolor}
\usepackage{academicons}
\definecolor{orcidlogocol}{HTML}{A6CE39}

\usepackage[pagewise]{lineno}

\usepackage{changes}

\yil{}
\vol{}
\fpage{}
\lpage{}
\doi{}

\title{Weak deflection angle of black-bounce traversable wormholes using Gauss-Bonnet theorem in the dark matter medium }

\author[AL\.{I} \"{O}VG\"{U}N]{
\textbf{AL\.{I} \"{O}VG\"{U}N$^{1}$\thanks{ali.ovgun@emu.edu.tr}~\href{https://orcid.org/0000-0002-9889-342X}{}}\\
\\ 
$^{1}$Physics Department, Eastern Mediterranean
University, Famagusta, North Cyprus \\ 99628 via Mersin 10, Turkey
\\ [1.8em]

\rec{.201}
\acc{.201}
\finv{..201}
}

\input{fizik.tex}

\begin{document}

\maketitle

\begin{abstract}In this paper, we first use the optical metrics of black-bounce traversable wormholes to calculate the Gaussian curvature. Then we use the Gauss-Bonnet theorem to obtain the weak deflection angle of light from the black-bounce traversable wormholes. Then we investigate the effect of dark matter medium on weak deflection angle using the Gauss-Bonnet theorem. We show how weak deflection angle of wormhole is affected by the bounce parameter $a$. Using the Gauss-bonnet theorem for calculating weak deflection angle shows us that light bending can be thought as a global and topological effect. 

\keywords{Relativity and gravitation; Gravitational lensing; Deflection angle; Wormholes; Gauss-Bonnet theorem.}
\end{abstract}

\section{Introduction}

Root of gravitational lensing is the deflection of light by gravitational fields such as a planet, a black hole, or dark matter predicted by Einstein's general relativity, in the weak-field limit \cite{Bartelmann:1999yn,Bozza:2009yw}. Weak deflection is used to detect dark matter filaments, and it is important topic because it helps to understand the large-scale structure of the universe \cite{Bartelmann:1999yn,Keeton:1997by}.

One of the important method to calculate the weak deflection angle using optical geometry is proposed by Gibbons and Werner (GW), which is known as Gauss-Bonnet theorem (GBT) \cite{Gibbons:2008rj,Werner:2012rc}. In the method of GW, deflection angle is considered as a partially topological effect and can be found by integrating the Gaussian optical curvature of the black hole space using:
 \cite{Gibbons:2008rj}
\begin{equation}
\hat{\alpha}=-\int\int_{D_{\infty}}K\mathrm{d}S,
\end{equation}
where $\hat{\alpha}$ is a deflection angle, $K$ is a Gaussian optical curvature, $dS$ is an optical surface and the $D_{\infty}$ stands for the infinite
domain bounded by the light ray, excluding the lens. Since the GW method provides a unique perspective, it has been applied to various types of black hole spacetime or wormhole spacetime \cite{Takizawa:2020dja}-\cite{Ono:2018ybw}.

In this paper, our main motivation is to explore weak deflection angle of black-bounce traversable wormholes \cite{Simpson:2019cer,Nascimento:2020ime} using the GBT and then extend our motivation of this research is to shed light on the effect of dark matter medium on the weak deflection angle of black-bounce traversable wormhole using the GBT. Note that the refractive index of the medium is supposed that it is spatially non-uniform but one can consider that it is uniform at large distances \cite{Latimer:2013rja}-\cite{Er:2013efa}. To do so, the photons are thought that may be deflected through dark matter due to the dispersive effects, where the index of refractive $n(\omega)$ which is for the scattering amplitude of the light and dark-matter in the forward.

\section{Calculation of weak deflection angle from black-bounce traversable wormholes
using the Gauss-Bonnet Theorem}

The ``black-bounce form''\ for the spacetime metric of traversable
wormhole is: \cite{Simpson:2019cer}
\begin{equation}d s^{2}=-\left(1-\frac{2 M}{\sqrt{r^{2}+a^{2}}}\right) d t^{2}+\left(1-\frac{2 M}{\sqrt{r^{2}+a^{2}}}\right)^{-1} d r^{2}+\left(r^{2}+a^{2}\right)\left(d \theta^{2}+\sin ^{2} \theta d \phi^{2}\right).\end{equation}

It is noted that the parameter $a$ stands for the bounce length scale and when $a=0$, it reduces to the Schwarzschild solution. We restrict ourselves to the equatorial coordinate plane ($\theta=\frac{\pi}{2}$),
so that the black-bounce traversable wormhole spacetime becomes 
\begin{equation}
ds^{2}=-\left(1-\frac{2 M}{\sqrt{r^{2}+a^{2}}}\right)dt^{2}+\left(1-\frac{2 M}{\sqrt{r^{2}+a^{2}}}\right)^{-1}{dr^{2}+\left(r^{2}+a^{2}\right)d\phi^{2}}.
\end{equation}
Then the optical geometry of the black-bounce traversable wormhole spacetime is found by using

\begin{equation}
g_{^{opt}\alpha\beta}=\frac{g_{\alpha\beta}}{-g_{00}},
\end{equation}
\begin{equation}
\mathrm{d}t^{2}=\frac{\mathrm{d}r^{2}}{\left(1-\frac{2 M}{\sqrt{r^{2}+a^{2}}}\right)^2}+\frac{(r^{2}+a^2)\mathrm{d}\varphi^{2}}{\left(1-\frac{2 M}{\sqrt{r^{2}+a^{2}}}\right)}.
\end{equation}
The Gaussian optical curvature $K$ for the black-bounce traversable wormhole optical space is calculated as
follows: 
\begin{equation}
K\simeq\\3\,{\frac {{m}^{2}}{{r}^{4}}}-2\,{\frac {m}{{r}^{3}}}-15\,{\frac {{a}^
{2}{m}^{2}}{{r}^{6}}}+10\,{\frac {{a}^{2}m}{{r}^{5}}}-{\frac {{a}^{2}
}{{r}^{4}}}. \label{curvature}
\end{equation}

Then we should use the Gaussian optical curvature in GBT to find deflection angle because the GBT is a theory which links the   intrinsic geometry of the
2 dimensional space with its topology ($D_{R}$ in $M$, with boundary
$\partial D_{R}=\gamma_{\tilde{g}}\cup C_{R}$) \cite{Gibbons:2008rj}: 
\begin{equation}
\int\limits _{D_{R}}K\,\mathrm{d}S+\oint\limits _{\partial D_{R}}\kappa\,\mathrm{d}t+\sum_{i}\epsilon_{i}=2\pi\chi(D_{R}),
\end{equation}
in which $\kappa$ is defined as the geodesic curvature ($\kappa=\tilde{g}\,(\nabla_{\dot{\gamma}}\dot{\gamma},\ddot{\gamma})$),
so that $\tilde{g}(\dot{\gamma},\dot{\gamma})=1$. It is noted that $\ddot{\gamma}$ is an unit
acceleration vector and $\epsilon_{i}$ is for the exterior angles at the $i^{th}$ vertex. 
Jump angles are obtained as $\pi/2$ for $r\rightarrow\infty$. Then we find that $\theta_{O}+\theta_{S}\rightarrow\pi$.
If $D_{R}$ is a non-singular, the Euler characteristic becomes $\chi(D_{R})=1$, hence GBT becomes
\begin{equation}
\iint\limits _{D_{R}}K\,\mathrm{d}S+\oint\limits _{\partial D_{R}}\kappa\,\mathrm{d}t+\theta_{i}=2\pi\chi(D_{R}).\label{gaussbonnet}
\end{equation}

The Euler characteristic number $\chi$ is 1, then the remaining
part yields $\kappa(C_{R})=|\nabla_{\dot{C}_{R}}\dot{C}_{R}|$ as
$r\rightarrow\infty$. The radial component of the geodesic curvature
is given by 
\begin{equation}
\left(\nabla_{\dot{C}_{R}}\dot{C}_{R}\right)^{r}=\dot{C}_{R}^{\varphi}\,\partial_{\varphi}\dot{C}_{R}^{r}+\Gamma_{\varphi\varphi}^{r}\left(\dot{C}_{R}^{\varphi}\right)^{2}.\label{izo1}
\end{equation}
For large limits of $R$, $C_{R}:=r(\varphi)=r=const.$ we obtain
\begin{equation}
\left(\nabla_{\dot{C}_{R}^{r}}\dot{C}_{R}^{r}\right)^{r}\rightarrow-\frac{1}{r},
\end{equation}
so that $\kappa(C_{R})\rightarrow r^{-1}$. After that it is not hard to see that  $\mathrm{d}t=r\,\mathrm{d}\,\varphi$, where $\kappa(C_{R})\mathrm{d}t=\mathrm{d}\,\varphi.$ The GBT reduces to this form
\begin{equation}
\iint\limits _{D_{R}}K\,\mathrm{d}S+\oint\limits _{C_{R}}\kappa\,\mathrm{d}t\overset{{r\rightarrow\infty}}{=}\iint\limits _{S_{\infty}}K\,\mathrm{d}S+\int\limits _{0}^{\pi+\hat{\alpha}}\mathrm{d}\varphi.
\end{equation}

The light ray follows the straight line so that, we can assume that $r_{\gamma}=b/\sin\varphi$ at zeroth order. The weak deflection angle can be calculated using the formula:
\begin{equation}
\hat{\alpha}=-\int\limits _{0}^{\pi}\int\limits _{r_{\gamma}}^{\infty}K\,r\,\mathrm{d}r\,\mathrm{d}\varphi.\label{angle}
\end{equation}

Using the Gaussian optical curvature \eqref{curvature}, we calculate the weak deflection angle of black-bounce traversable wormholes up to second order terms:

\begin{eqnarray}
\hat{\alpha} & \approx &  4\,{\frac{m}{b}}+{\frac {{a}^{2}\pi}{4{b}^{2}}}.\label{myi1}
\end{eqnarray}

Note that it is in well agreement with the \cite{Nascimento:2020ime} in leading order terms.

\section{Deflection angles of photon through dark matter medium from black-bounce traversable
wormholes}

In this section, we investigate the effect of dark matter medium on the weak deflection angle. To do so, we use the refractive index for the dark matter medium \cite{Latimer:2013rja}:

\begin{equation}n ( \omega ) =1+\beta A_0+A_2 \omega^2. \label{dm} \end{equation}

Note that $\beta=\frac { \rho_0 } { 4 m ^ { 2 } \omega ^ { 2 } } $, $\rho_0$ is the mass density of the scattered dark matter particles,  $A _ { 0 } = - 2 \varepsilon ^ { 2 } e ^ { 2 } \text { and } A _ { 2 j } \geq 0
$. The terms in $
\mathcal { O } \left( \omega ^ { 2 } \right)
$ and higher terms are related to the polarizability of the dark-matter candidate. 

 The order of $\omega^{-2}$ is for the charged dark matter candidate and $\omega^{2}$ is for a neutral dark matter candidate. In addition, the linear term in $\omega$ occurs when parity and charge-parity asymmetries are present. The 2 dimensional optical geometry of the wormhole is: 
\begin{equation}
d\sigma^{2}=n^{2}\bigg(\frac{dr^{2}}{\left(1-\frac{2 M}{\sqrt{r^{2}+a^{2}}}\right)^2 }+\frac{(r^{2}+a^2)}{\left(1-\frac{2 M}{\sqrt{r^{2}+a^{2}}}\right) }d\varphi^{2}\bigg),\label{eq:opt}
\end{equation}

and 
\begin{equation}
\frac{d\sigma}{d\varphi}\bigg|_{C_{R}}=n\bigg(\frac{r^{2}+a^2}{\left(1-\frac{2 M}{\sqrt{r^{2}+a^{2}}}\right)}\bigg)^{1/2}.
\end{equation}

Using the GBT within optical geometry of black-bounce traversable wormhole, we obtain the weak deflection angle in a dark matter medium:
\begin{equation}
\lim_{R\rightarrow\infty}\int_{0}^{\pi+\alpha}\left[\kappa_{g}\frac{d\sigma}{d\varphi}\right]\bigg|_{C_{R}}d\varphi=\pi-\lim_{R\rightarrow\infty}\int\int_{D_{R}}\mathcal{K}dS.\label{alpha-bonnet}
\end{equation}
First we calculate the Gaussian optical curvature at linear order of $M$:
\begin{equation}
\mathcal{K}\thickapprox 
10\,{\frac {{\it csgn} \left( r \right) {a}^{2}M}{ \left( {\it A_2}\,{
\omega}^{2}+\beta\,{\it A_0}+1 \right) ^{2}{r}^{5}}}-2\,{\frac {{\it 
csgn} \left( r \right) M}{ \left( {\it A_2}\,{\omega}^{2}+\beta\,{\it 
A_0}+1 \right) ^{2}{r}^{3}}}-{\frac {{a}^{2}}{ \left( {\it A_2}\,{\omega
}^{2}+\beta\,{\it A_0}+1 \right) ^{2}{r}^{4}}}.\label{mK}
\end{equation}

After that we find
\begin{equation}
\lim_{R\rightarrow\infty}\kappa_{g}\frac{d\sigma}{d\varphi}\bigg|_{C_{R}}=1.\label{khomons}
\end{equation}

Then for the limit of $R\rightarrow\infty$, the deflection angle in dark matter medium can be calculated using the GBT as follows:

\begin{equation}
\alpha=-\lim_{R\rightarrow\infty}\int_{0}^{\pi}\int_{\frac{b}{\sin\varphi}}^{R}\mathcal{K}dS.\label{alpha1}
\end{equation}
Hence, we obtain the weak deflection angle in dark matter medium as follows: 
\begin{equation}
\alpha=4\,{\frac {M}{b
\Psi }}+{\frac {{a}^{2}\pi}{4{b}^{2}  \Psi }},
\label{phs}
\end{equation}
where
\begin{equation}
   \Psi= {{\it A_2}}^{2}{\omega}^{4}+2\,
{\it A_0}\,{\it A_2}\,\beta\,{\omega}^{2}+{{\it A_0}}^{2}{\beta}^{2}+2\,{
\it A_2}\,{\omega}^{2}+2\,\beta\,{\it A_0}+1,
\end{equation}
which agrees with the known expression found using another method.
Of course, in the absence of the dark matter medium ($\Psi=0$),
this expression reduces to the known vacuum formula $\alpha\approx 4\,{\frac{m}{b}}+{\frac {{a}^{2}\pi}{4{b}^{2}}}.$ Hence, we find that the deflected photon through the dark matter around the black-bounce traversable wormhole has large deflection angle compared to black-bounce traversable wormhole without dark matter medium.

\section{Conclusion}

In this paper, we have studied the weak deflection angle of black-bounce traversable wormholes using the GBT. Then we have investigated the effect of dark matter medium on the weak deflection angle of black-bounce traversable wormholes. Note that refractive index is taken spatially non-uniform, and it is uniform
at large distances. Hence it is concluded that the deflection angle by black-bounce traversable wormholes increases with increasing the bounce parameter $a$, on the other hand the deflection angle decreases in a increasing medium of dark matter. It is showed that how weak deflection angle of wormhole is affected by the bounce parameter $a$. Moreover we use the Gauss-bonnet theorem for calculating weak deflection angle which proves us that light bending can be thought as a global and topological effect.

\end{document}

%% file: fizik.tex
\newcommand{\bc}{\begin{center}}
\newcommand{\ec}{\end{center}}

\numberwithin{equation}{section}

\renewcommand{\phi}{\varphi}